\documentclass[preprint, showpacs,amsmath,amssymb]{revtex4}
\usepackage{graphicx} %include figure files
\usepackage{dcolumn}%alingn table columns on decimal point
\usepackage{bm}%bold math
\begin{document}
\title{Spin-dependent transmission in waveguides with periodically modulated
strength of the spin-orbit interaction}
\author{ X. F. Wang$^\dagger$ and
P. Vasilopoulos$^\star$ \\
\ \\}
\address{ Department of Physics; Concordia University\\
 1455 de Maisonneuve Ouest \\ Montr\'{e}al, Qu\'{e}bec,   H3G  1M8,  Canada\\
 \ \\  }
%\maketitle
\date{\today}
\begin{abstract}
The electron transmission $T$ is evaluated through waveguides, in
which the strength of the spin-orbit interaction(SOI) $\alpha$ is
varied periodically, using the transfer-matrix technique. It is
shown that $T$ exhibits a {\it spin-transistor} action,  as a
function of $\alpha$ or of the length of one of the two subunits
of the unit cell, provided only one mode is allowed to propagate
in the waveguide. A similar but not periodic behavior occurs as a
function of the incident electron energy. A transparent formula
for $T$ through one unit is obtained and helps explain its
periodic behavior. The structure considered is a good candidate
for the establishment of a realistic spin transistor.
\end{abstract}


\maketitle

Going beyond the speed limit of conventional electronic digital devices
has motivated the rapidly increasing interest in electronic spin-based devices,
which may promisingly be operated at  much higher
speeds and less energy consumption. Consequently, various mechanisms
of realizing the manipulation of the spin in nanoscale devices are under active
investigation. One  approach is employing the spin-orbit interaction
(SOI). As a linear term of the SOI in semiconductor nanostructures,
the Rashba coupling has attracted considerable attention
experimentally and theoretically in the past years. Originated from studies of
the electron
dynamics in a crystal electric field, the Rashba coupling works as a local effective
magnetic field perpendicular to both the electronic momentum and
the electric field. As a result, the energy degeneracy of
spins is lifted and the electronic spin precesses
under this coupling. An
electron gas with specified momentum and energy
can then be used as a filter for spins, through which electrons of any spin
orientation can be selected from randomly spin-polarized (or
unpolarized) electron gases \cite{vali}. 
A spin transistor has been proposed in waveguides which control of  the spin-polarized electronic flux by exploiting the spin precession due to the SOI \cite{datt} or by combining it with the modulation of the transmission resulting from attaching stubs to the waveguides \cite{wang}.

In  previous work \cite{wang} we studied  ballistic transport and
spin-transistor behavior in stubbed waveguides in the presence of SOI
and found encouraging results, among others a nearly {\it square-wave} form of the transmission as a function of some stub parameters. From an experimental point of view, however, it may be very difficult   to construct controllable stubs periodically attached to a quantum wire. Recent experiments  show that the Rashba coupling strength can be well
controlled by applying a bias \cite{nit} or
adjusted with the help of band engineering \cite{kog2}.
In this paper we show that we can  control the spin flux through  waveguides, without stubs, with periodically modulated SOI strength.  The details are as follows.

\begin{figure}[h]
\vspace{-3.5cm}
\includegraphics*[width=100mm,height=100mm]{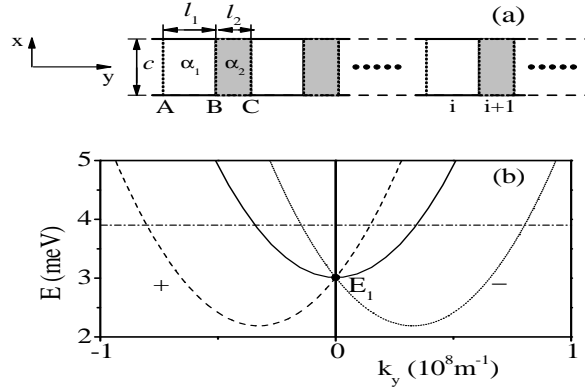}
\vspace{-1.5cm}
\caption{(a) Schematics of a waveguide, of width $c$, with
periodically modulated strength of the SOI. Within one unit $l_1,
l_2$ and $\alpha_1, \alpha_2$ are the lengths and SOI strengths of
the subunits AB and BC, respectively. (b) Dispersion relation
for a waveguide. The dashed and dotted curves show the + and -
branches for finite  strength $\alpha$; the solid curve is for
$\alpha=0$. The intersection of all curves (solid circle) denotes the energy $E_1$ of the lowest subband
due to the confinement along the $x$ axis.}
\label{fig1}
\end{figure}

In the absence of a magnetic field the spin degeneracy of the
quasi one-dimensional electron gas
(Q1DEG) energy bands at ${\bf k}\ne 0$ is lifted by the coupling of
the electron spin with its orbital motion. This coupling is
described by the Rashba Hamiltonian
 \begin{equation}
 H_{so}=\alpha (\vec{ \sigma }\times
 \vec p)_z/ \hbar=i\alpha [\sigma _y \partial
 /\partial x-\sigma _x \partial/\partial y)].
 \end{equation}
 Here the waveguide is along the $y$ axis and the confinement along the $x$ axis,
 cf. Fig. \ref{fig1}.  The parameter $\alpha$  measures the strength of the coupling;
 $\vec{\sigma}=(\sigma_x,\sigma_y,\sigma_z)$  denotes
 the spin Pauli matrices,  and $ {\vec p}$ is the momentum
 operator.

We treat  $H_{so}$ as a perturbation. With
$\Psi =| n, k_{y},\sigma\rangle
=e^{ik_{y}y}\phi_{n}(x)|\sigma\rangle$  the
eigenstate in each region in Fig. \ref{fig1}(a) the
unperturbed states satisfy  $H^{0}|n, k_y, \sigma\rangle=E_{n}^{0}|n, k_y, \sigma\rangle$ with
$E_{n}^{0}=E_{n}+\lambda k_{y}^{2}$, $\lambda=\hbar^2/2m^*$, and
$\phi_{n}(x)$  obeys $[-\lambda d^2/dx^2+V(x)]\phi_n(x)=E_n\phi_n(x)$,
where $V(x)$ is the confining potential assumed to be square-type and
high enough that  $\phi_n(x)$ vanishes at the boundaries.
The perturbed ($H_{so}\neq 0)$ eigenfunction,
is written as $\sum_{n,\sigma}A^{\sigma}_n\phi_n(x)|\sigma\rangle$. $H_{so}$ is
a $2\times 2$ matrix. Combining it with
the  $2\times 2$ diagonal
matrix $H^{0}$ and using  $H\Psi=(H^{0}+H_{so})\Psi=E\Psi$ leads to the equation
 %Fig. 1
%Eq.()
\begin{equation}
\left[
\begin{array}{cc}
E_{n}^0 -E & \alpha k_{y}\\
\alpha k_{y} & E_{n}^0-E
\end{array}
\right]
\left(
\begin{array}{c}
A_{n}^{+}\\
A_{n}^{-}
\end{array}
\right)
+\alpha\sum_m
J_{nm}
\left[
\begin{array}{cc}
0 & 1 \\
-1 & 0
\end{array}
\right]
\left(
\begin{array}{c}
A_{m}^{+} \\
A_{m}^{-}
\end{array}\right)
=0.
\label{cross}
\end{equation}

The index $n$ labels the discrete subbands resulting from the confining
potential $V(x)$. If the subband mixing is negligible, we can take $J\approx 0$ in Eq. (\ref{cross});
 the resulting eigenvalues $E\equiv E^{\pm}(k_y)$, plotted in Fig. \ref{fig1}, read
%Eq.()
 \begin{equation}
 E^{\pm}(k_y)=E_n+\lambda k_y^2\pm\alpha k_y.
 \label{disp}
 \end{equation}
 The eigenvectors corresponding to $ E^{+}, E^{-}$ satisfy $A_{n}^\pm
 =\pm A^\mp_{n}$. Accordingly, the spin eigenfunctions are taken as
$|\pm\rangle={\tiny\left(\array{c} 1 \\ \pm 1 \endarray \right)}/\sqrt{2}$.
For the same energy the difference in wave vectors $k^+_{y}$ and $k^-_{y}$ for
the two spin orientations is
 \begin{equation}
 k^-_{y}-k^+_{y}=2m^*\alpha/\hbar^2= \delta.
 \label{difk}
 \end{equation}
The dispersion relation $E^\pm(k_y)$ vs $k_y$ resulting from Eq.
(\ref{disp})
is shown in Fig. \ref{fig1}. For the same energy $E$,
shown by the  dash-dotted line in Fig. 1(b),  there are four $k_y$ values
and a phase shift $\delta$ between the positive or negative $k^+_y$ and
$k^-_y$ values of the branches $E^+$ and $E^-$.

The procedure outlined above applies to all waveguide segments in
Fig. \ref{fig1}. When the electron  energy is low and only one
subband is occupied, we omit the subband label $n$ from the
wave function and use the segment label $i$ instead. Then the
eigenfunction $\phi_i$ of energy $E$ in   waveguide segment $i$
reads
%Eq.()
\begin{equation}
\phi_i (x,y)=\sum_{\pm} \left[c_i^{\pm}e^{ik^{\pm}_y y}|\pm\rangle+
\bar{c}_i^{\pm}e^{-ik^{\mp}_y y}|\pm\rangle\right]
\sin[(x+c/2)\pi/c].
\label{wavf}
\end{equation}

We now match the wave function and its flux at the interfaces between  the $i$ and  $i+1$ segments. We do so,
in line with Ref. \onlinecite{mole}, because in  transport
through materials with different Rashba parameters, the continuity
of the derivative of the wave function may not hold 
and is not clear how to
modify it appropriately. The velocity operator is given by
\begin{equation}
\hat{v}_y=\frac{\partial H}{\partial p_y}
=\left[
\begin{array}{cc}
-i\frac{\hbar}{m^*}\frac{\partial }{\partial y}&\frac{\alpha}{\hbar} \\
\frac{\alpha}{\hbar}&-i\frac{\hbar}{m^*}\frac{\partial }{\partial y}
 \end{array}
\right].
\end{equation}
The continuity of the wave function at the interface $y=y_{i, i+1}$, between
the $i$ and  $i+1$ segments, gives $\phi_{i+1} (x,y_{i, i+1})=\phi_i (x,y_{i, i+1})$ and that of the flux
$\hat{v}_y \phi_{i+1}(x,y)|_{y_{i, i+1}}=\hat{v}_y \phi_{i}(x,y)|_{y_{i, i+1}}$. This
connects the coefficients $c_i$ and $c_{i+1}$ of
the $i$ and $i+1$ segments in the manner

%Eq.()
\begin{equation}
\hat{Q}_{i}\left(
\begin{array}{c}
c^{\pm}_{i} \\
\bar{c}^{\pm}_{i}
\end{array}
\right)
=\hat{Q}_{i+1}\left(
\begin{array}{c}
c^{\pm}_{i+1} \\
\bar{c}^{\pm}_{i+1}
\end{array}
\right), \quad\quad
%\end{equation*}
%where
%\begin{equation*}
\hat{Q}_{i}=\left[
\begin{array}{cc}
1 & 1 \\
\Delta_{i} & -\Delta_{i}
\end{array}
\right],
\end{equation}
where $\Delta_{i}=[m^{\ast 2}\alpha_{i}^{2}+2m^{\ast}(E-E_1)]^{1/2}$.
One noteworthy feature
here is the independence of $\Delta_{i}$ on the sign of $\alpha
_{i }$ or  direction of the Rashba electric field. If electrons pass through
an interface connecting two waveguide segments with the same $\alpha$ but
opposite orientations, the electrons in different branches simply
exchange their momenta but suffer no reflection.

The total transfer matrix $\hat{M}$ of a series of waveguide segments, cf. Fig. \ref{fig1}, for
electrons in the $\pm $ branch is given by

\begin{equation}
\hat{M}^{\pm }=\prod_{i=1,n}(\hat{P}_{i}^{\pm }\hat{Q}^{-1}_{i}\hat{Q}_{i+1}), \quad\quad
%\end{equation*}
%with the transfer matrix of segment $i$%
%\begin{equation*}
\hat{P}_{i}^{\pm }=\left[
\begin{array}{cc}
e^{-ik_{y}^{\pm }l_i} & 0 \\
0 & e^{ik_{y}^{\mp }l_i}
\end{array}
\right],
\end{equation}
with $\hat{P}_i^{\pm }$ being the transfer matrix through segment $i$.

\begin{figure}[h]
\vspace{-3.5cm}
\includegraphics*[width=130mm,height=120mm]{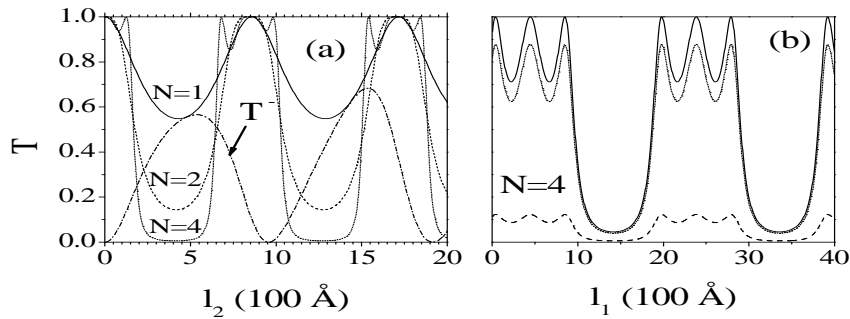}
\vspace{-3cm}
\caption{(a) Transmission versus length $l_2$ with fixed $l_1=1050$ \AA.
The dash-dotted curve shows the spin-down transmission $T^-$ for $N=1$, the other curves show the {\it total} transmission.
(b) The total, spin-up, and spin-down transmissions are shown, respectively, by the solid, dotted, and dahed curve versus the length $l_1$ for fixed $l_2=1050$ \AA.  $N$ is the number of units and $E=3.2$ meV.}
\label{fig2}
\end{figure}

In a waveguide with only one segment,  of strength
$\alpha_{2}$ and length $l_2$, sandwiched between two segments of strength
  $\alpha_{1}$, the transmission is obtained as
\begin{equation}
T=\frac{1}
{\cos^{2}(\Delta_{2}l_2)+r
\sin^{2}(\Delta_{2}l_2)},
\label{trans}
\end{equation}
\noindent
where $r=(\Delta^2_1+\Delta^2_2)^2/4\Delta^2_1\Delta^2_2$. This is a sinusoidal dependence with a maximum
$T_{max}=1$ for $\sin(\Delta_2 l_2)=0$ and a minimum $T_{min}=1/r$.
The most efficient modulation of the transmission is obtained if
we increase the difference between $\alpha_1$ and $\alpha_2$ and minimize the energy difference 
$E-E_1$. In a waveguide, such as the unit ABC shown in Fig. \ref{fig1}(a),
made of InGaAs with $\alpha_1=0$, $\alpha_2=5\times 10^{-11}$ eV m, and $E-E_1=0.2$ meV,
the transmission oscillates between 1 and 0.55 with a period $l_2= 858$ \AA.
As in a simple waveguide of length $L$, the spin orientation is determined
by the total phase difference $\theta=\delta_1(L-l_2)+\delta_2 l_2$, acquired by
electrons in different
branches  during the propagation with $\delta_1, \delta_2$ determined from Eq. (\ref{difk}).
If only spin-up electrons are incident,
the spin-up transmission will be $T^+=T \cos^2(\theta/2)$ and the
spin-down one $T^-=T \sin^2(\theta/2)$. With the above parameters, the spin of electrons will flip
periodically with a period $l_2=958$ \AA. As
Eq. (\ref{trans}) makes clear,
in which $l_2$ appears only in the factors $\cos(...)$ and $\sin(...)$, the transmission shows a periodic dependence on
$l_2$ with well-defined peaks and gaps. The total (spin-down) transmission $T$ ($T^-$) resulting from Eq. (\ref{trans})
for $N=1$ is shown by the solid (dash-dotted) curve in Fig. \ref{fig2}(a).
%Fig. 2

In the following we consider the transmission
through a waveguide, consisting of $N$ identical units, with periodically modulated $\alpha$.
For the sake of convenience we assume that the  incident electrons are  spin-up polarized. If they are spin-down polarized, the only change occurs in the phase $\theta$, the ratio of the spin-up to spin-down contributions changes but  the results for the total transmission remain unaffected.
As shown in Fig. \ref{fig1}(a), each unit is identical to the one labelled ABC and consists of two segments
$i$ and $i+1$ of length
$l_{i}=l_1$ (AB) and $l_{i+1}=l_2$ (BC) with strengths $\alpha_{i}=\alpha_1$ and $\alpha_{i+1}=\alpha_2$,
respectively.
If not otherwise specified, the parameters $c=500$ \AA, $\alpha_1=0$, and zero temperature are used.
The dependence of the total  transmission on the length $l_2$ for $N=2, 4$  units is shown by the dashed and dotted curve, respectively, in Fig. \ref{fig2}(a), together with the result for $N=1$ commented above. As can be seen, with
increasing $N$ the dips in the transmission become deeper and
smaller-amplitude dips appear at the main peaks. If $l_2$ is
fixed, as shown in Fig. \ref{fig2}(b) where $N=4$, the transmission shows similar peaks and dips  as function
of $l_1$ when more than one units are considered.
Since $\alpha_1=0$ the spin polarization does not change and the percentage of the spin-up (spin-down) remains constant as $l_1$ varies. However, this percentage is very sensitive to the choice of $l_2$ as perusal of Fig. 2(a) shows.

\begin{figure}[h]
\vspace{-1.5 cm}
\includegraphics*[width=80mm,height=70mm]{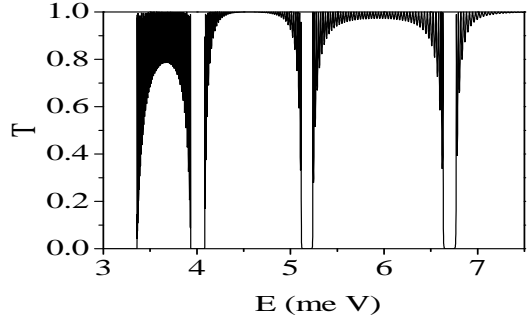}
\vspace{-1cm}
\caption{Transmission $T$ as a function of the energy for a waveguide of $N=50$ units with $l_1=l_2=1050$ \AA \ and
 $\alpha_2=6\times 10^{-11}$ eV m.}
\label{fig3}
\end{figure}

The total transmission $T$, at zero temperature, as a function of
the electron energy $E$ through a  waveguide
composed of 50 identical units is shown in Fig. \ref{fig3}. Each unit consists of two segments of length
$l_1=l_2=1050$ \AA \ with SOI strength $\alpha_2=6\times 10^{-11}$ eV m.
Transmission gaps appear when the energy $E$ is below $3.35$ meV, near 4 meV, 5.2 meV, and 6.7 meV.
As is usually the case in periodically modulated
nanostructures, the transmission bands appear as
a set of oscillations between the gaps. If the incident electrons are spin-polarized, the spin-up or spin-down
transmission oscillate with increasing energy under the envelop of the total transmission.
 %Fig. 3

\begin{figure}[h]
\vspace{-5cm}
\includegraphics*[width=120mm,height=160mm]{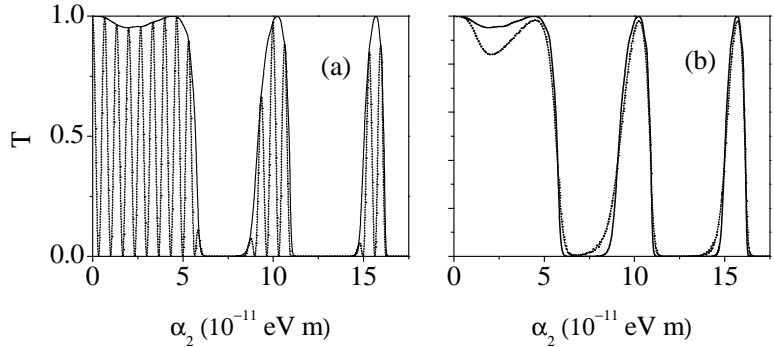}
\vspace{-5.5cm}
\caption{Transmission as a function of the strength $\alpha_2$,
at temperature $T=0.2$ K,
with $l_1=l_2=900$ \AA, N=8, and $E_F=3.3$ meV.
(a) The solid (dotted) curve is the {\it total} (spin-up) transmission. (b) {\it Total} transmission versus strength $\alpha_2$ for $T=0.2$ K
(solid curve) and  $T=0.5$ K (dotted curve).}
\label{fig4}
\end{figure}

As usual, by integrating the  zero-temperature transmission over the energy \cite{wang} we obtain the finite-temperature
one and see that raising the temperature
rounds off the zero-temperature transmission profile. In Fig. \ref{fig4}(a)
we show the transmission as a function of the  strength $\alpha_2$ in one segment at temperature $T=0.2$ K.  The parameters used are given in the caption and the incident electrons are spin-up polarized.
The spin-up transmission $T^+$ is shown by the
dotted curve; the spin-down transmission $T^-$, not shown, has the
same oscillating dependence on $\alpha_2$ as $T^+$ but with opposite phase.
The total transmission, shown by the solid line, serves as the envelop function of the rapidly oscillating $T^+$ and
$T^-$ transmissions and shows two wide gaps below $\alpha_2=15\times 10^{-11}$ eV m.
These gaps have a well-defined, approximately square-wave form at this temperature. With increasing
temperature, the square-wave gaps of $T$ become round, narrower, and shallower. This is seen in
Fig. \ref{fig4}(b) where
we replot the total transmission for $T=0.2$ K (solid curve), taken from Fig. \ref{fig4}(a),
and for $T=0.5$ K (dotted curve).

%Fig. 4

All the results presented here are valid when only {\it one} mode propagates in the waveguide.
If more modes allowed, they become somewhat more complicated but remain qualitatively the same; details will be given
elsewhere. We notice though that the present modulation and approximate square-wave form of the transmission are
obtained by varying periodically the
SOI strength $\alpha$ but without attaching stubs to the waveguide. Accordingly, some of
the constraints of our earlier work \cite{wang} are
considerably relaxed, experimentally speaking. Also, relative to this work here the width of the waveguide is about a factor of 2 larger
and, due to the periodic dependence of the  transmission on the width $l_2$, there is no restriction
on the latter whereas previously the width of the stubs had to be within certain limits. All these features
should facilitate
the fabrication of the relevant samples. The latter could be prepared by a spatially {\it periodic} application
of gates which control the value of $\alpha$.

An  interesting feature of the results is that in waveguides with the same strength $\alpha$ everywhere \cite{datt} the {\it total} transmission is always equal to 1 whereas here it is not, see Eq. (9),
the comments that follow,  and Fig. 2(a). This
means that changing $\alpha$ is equivalent to introducing barriers to the  electron transmission. However, the spin
orientation is determined by the total phase difference $\theta=\delta_1(L-l_2)+\delta_2 l_2$, as in a simple waveguide, cf. Eq. (4).

A relative improvement of the results presented here for a waveguide of $N$ simple identical can be obtained if we
consider a waveguide of $N$ identical but composite units, i.e., units with further structure in the manner of
Ref. \onlinecite{wang1}. In this way additional barriers are introduced and further control the electron transmission.
Details will be given elsewhere.

In summary, we showed that the transmission in waveguides with periodically modulated SOI
strength $\alpha$ can have an approximate {\it square-wave} profile as a function  of $\alpha$ or of the length of one
of the two subunits  of the unit cell, provided only one mode is allowed to propagate in the waveguide.


\begin{references}
\bibitem{vali} M. Valin-Rodriguez, A. Puente, and L. Serra, cond-mat/0211694.

\bibitem{datt}  S. Datta and B. Das, Appl. Phys. Lett. {\bf 56}, 665 (1990).

\bibitem{wang} X. F. Wang, P. Vasilopoulos, and F. M. Peeters, Appl. Phys. Lett. {\bf 80}, 1400 (2002); Phys. Rev. B {\bf 65}, 165217 (2002).

\bibitem{nit} J. Nitta, T. Akazaki, H. Takayanagi, T. Enoki, Phys. Rev.
Lett. {\bf 78}, 1335 (1997).

\bibitem{kog2} T. Koga, J. Nitta, T. Akazaki, and H. Takayanagi, Phys.
Rev. Lett. {\bf 89}, 046801 (2002).

\bibitem{mole}L. W. Molenkamp, G. Schmidt, and G. E. W. Bauer, Phys. Rev. B {\bf 64}, 121202 (2001).

\bibitem{wang1} X. F. Wang and P. Vasilopoulos,  Appl. Phys. Lett. {\bf 81}, 1636 (2002).

%\bibitem{poto}R. M. Potok, J. A. Folk, C. M. Marcus, and V. Umansky, Phys. Rev. Lett. {\bf 89}, 266602 (2002).
%
%\bibitem{mire} F. Mireles and G. Kirczenow, unpublished.
\end{references}
\end{document}